\documentclass[preprint]{aastex7}

\usepackage[utf8]{inputenc}
\usepackage[english]{babel}
\usepackage{natbib}
\usepackage{fp}
\usepackage[capbesideposition=right]{floatrow}
\usepackage{wrapfig}
\usepackage{mathtools}
\usepackage{amsmath}
\usepackage{supertabular,enumitem}
\usepackage{tabularx}
\usepackage{float}
\usepackage{graphicx}
\usepackage{multirow}

\usepackage{caption}
\usepackage{subcaption}

\usepackage{hyperref}

\shorttitle{Coronal Cells in Coronal Holes}
\shortauthors{Alzate et al.}
\accepted{August 3, 2025}
\graphicspath{{./}{figures/}}
\begin{document}

\title{Coronal Cells in Coronal Holes:\\Systematic Analysis and Implications for Coronal Evolution}

\correspondingauthor{Nathalia Alzate}

\author[orcid=0000-0001-5207-9628,gname=Nathalia,sname=Alzate]{Nathalia Alzate}
\affiliation{NASA Goddard Space Flight Center, Greenbelt, MD, USA}
\affiliation{ADNET Systems, Inc., Greenbelt, MD, USA}
\email[show]{nathalia.alzate@nasa.gov}

\author[orcid=0000-0001-6407-7574,gname=Simone,sname=Di Matteo]{Simone Di Matteo}
\affiliation{Physics Department, The Catholic University of America, Washington, DC, USA}
\affiliation{NASA Goddard Space Flight Center, Greenbelt, MD, USA}
\email{simone.dimatteo@nasa.gov}

\author[orcid=0000-0003-1380-8722,gname=Aleida,sname=Higginson]{Aleida Higginson}
\affiliation{NASA Goddard Space Flight Center, Greenbelt, MD, USA}
\email{aleida.k.higginson@nasa.gov}

\begin{abstract}

Using advanced processing techniques, we analyze high-cadence, high-resolution extreme ultraviolet images and show that, throughout the solar cycle, mid-latitude coronal holes (CHs) are made up of ubiquitous and space-filling funnel-shaped structures (or cells) anchored to unipolar magnetic flux concentrations in \added{network} lanes. We demonstrate that the coronal cells\added{,} previously documented in the magnetically closed regions\added{,} as well as coronal plumes\added{, inside CHs,} are a particular manifestation of \added{ubiquitous cells}. The cell properties depend on the magnetic field intensity at their footpoint and connectivity in the corona, either closing in opposite polarity regions (CF-cells) or extending to form open-field (\added{OF}-cells). \added{The \added{OF}-cells reach size scales on the order of super-granules and are characterized by dark lanes delimiting ray-like features both showing, at different levels, persistent jet-like ejections.} The cells’ lifetime mirrors that of magnetic flux concentrations revealed by magnetograms, slowly emerging and then disappearing in a matter of a few hours to a few days in a one-to-one correspondence. When a cell \added{forms} along the CH boundary, it can \added{alter the CH boundary by shrinking or expanding the CH by an approximately supergranular cell unit}. Therefore, coronal cells can contribute to the dynamics of CH \added{boundary}. We contextualize these observations in a ``coronal cell'' theory potentially able to provide an explanation for fine scale coronal structures and \added{jetting activity in polar coronal holes.}

%lead to \added{either} a CF-cell, shrinking the CH, or a \added{OF}-cell, expanding the CH \added{by an approximately supergranular cell unit. Therefore, coronal cells can contribute to the dynamics of CH and the amount of open magnetic flux on supergranular scales along the CH boundary.} We contextualize these observations in a ``coronal cell'' theory potentially able to provide explanation for fine scale coronal structures and \added{jetting activity in polar coronal holes.}

\end{abstract}



%------------------------------
\section{Introduction} 
\label{sec:intro}
%------------------------------

\begin{figure}[t!]
    \centering
    \includegraphics[width=\textwidth]{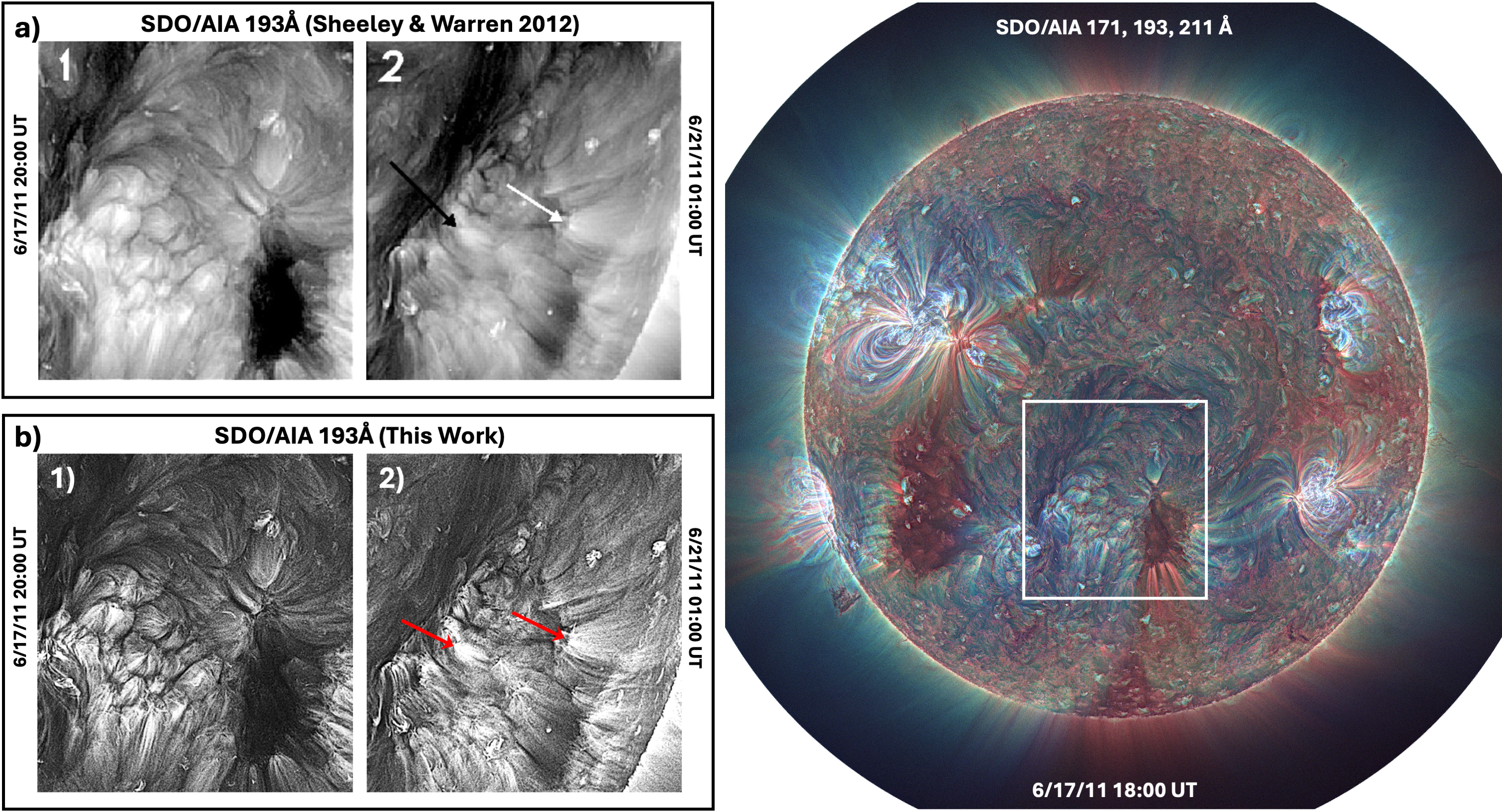}
    \caption{\footnotesize\textbf{Example of \added{the on disk and near limb view of the same} coronal cells in \added{original and processed} SDO/AIA images.} a) Adapted from \citet{sheeley2012}; b) Same but processed with the \added{Multiscale Gaussian Normalization technique} by \citet{Morgan2014}; Panels 1 in a) and b) are face-on views of a coronal cell network. Note that these were reported to be present only outside CHs. Panels 2 in a) and b) show the cells near the limb appearing as plumes (some indicated with arrows). c) Full disk 3-wavelength (171, 193 and 211 \AA) AIA composite of MGN processed images. The white rectangle indicates the region shown in panel a) and b).}
    \label{fig:sheeley-compare}
\end{figure}

Coronal cells were first discovered by \citet{sheeley2012} in the magnetically closed corona. These cells were reported in \added{Solar Dynamics Observatory/Atmospheric Imaging Assembly (SDO/AIA) Extreme Ultraviolet (EUV)} Fe XII 193 \AA\ observations with temperatures approximating 1.2 MK (Figure \ref{fig:sheeley-compare}a). Later studies noted the presence of sporadic cells in coronal holes (CHs) too (e. g., \citealt{sheeley2014}) visible not in  193 \AA\ but in 171 \AA\ observations, however, their presence in CHs has never been studied in detail. These types of structures, as originally documented in the magnetically closed corona, are the result of closely spaced radial flame-like structures that appear as cells when seen from above and as plumes when seen on the limb. Their bright centers are separated by narrow dark lanes with size of $\approx$30,000 km (supergranular size). Each of the coronal cells is centered above a concentration of photospheric flux in \added{Helioseismic Magnetic Imager (HMI)} magnetogram observations, which also display supergranular spatial scale sizes. These closed-field cells (CF-cells) are found most commonly between filament channels and CH boundaries, and have been subsequently used to derive information about the horizontal magnetic field in the associated filament channels \citep{sheeley2013,sheeley2014}.

Bright, relatively short-lived features (i. e., EUV plumes) were considered to occur only in CHs \citep{Poletto2015}. However, recent observations using AIA’s 171, 193, and 211 \AA\ EUV images and HMI magnetograms showed similar features in quasi-unipolar quiet, closed-field regions \citep{Wang2016, Avallone2018}. These features appear to be born, live, and die in synchrony with the compact majority-polarity magnetic flux concentration at their foot carried by converging supergranular convection flows \citep{Moore2023}. Consequently, these features become fairly discernible in AIA 171 \AA\ images when the majority-polarity network flux concentration starts to form. Once formed, the EUV plume luminosity, width, and upward extent remain roughly constant and maximal before fading to invisibility in AIA 171 \AA\ emission as photospheric convection flows break apart the network flux concentration. The EUV plumes are not observed in the 193 \AA\ passband (Fe XII emission from plasmas at temperatures around 1.6x10$^6$ K) and the 211 \AA\ passband (Fe XIV emission from plasmas at temperatures around 1x10$^6$ K) as clearly as in the 171 \AA\ passband (Fe IX emission from plasmas at temperatures around 6x10$^5$ K). These plumes exhibit similar structuring to the coronal cells, rising above magnetic flux concentrations, except that plumes are much shorter lived. In fact, coronal cells are more akin to ``network plumes'' which are a more diffuse and faint set of plumes (only observable through brightness integration over time) as opposed to visible short-lived ``beam plumes'' \citep{Gabriel2009}.

Small-scale and transient magnetic activity may play a role in the formation and evolution of plumes \citep{Raouafi2014}. Coronal jets prior to the formation of a plume were first described by \citet{Raouafi2008} and \citet{Raouafi2009} using \added{EUV and X-ray observations}. \citet{Raouafi2014} then suggested that during the main phase of a plume, small jets, or jetlets, and plume transient bright points occur ubiquitously at the plume footpoints throughout the lifetime of the plume. \citet{Kumar2022} found that these small-scale plasma outflows or jetlets, are produced at the base of several equatorial and polar plumes. According to their work, these jets then transition smoothly into bright plumelets that comprise the overlying plumes. Plumelets, according to \citet{Uritsky2021}, are filamentary substructures in plumes that may account for the brightness in the fully formed host plume. Additionally, \citet{Chitta2025} found that similar small-scale jet-like activity also occurs in the interplume region surrounding plumes.

Mid-latitude CHs are important for space weather, as they are the source of recurrent high speed streams responsible for a key subset of geomagnetic events. CH boundaries have also been hypothesized to be one possible source of the slow solar wind based on expansion factor and interchange reconnection models. Advanced image processing techniques (Figure \ref{fig:sheeley-compare}b panels) can reveal detailed structuring within CHs, similar to the ``candles on a cake'' in the quiet Sun corona, discovered by \citet{sheeley2012}. When examined in CHs, this detailed CH structuring may be directly connected to jets, plumes, and jetlets, recently hypothesized to be a source of the solar wind \citep{Uritsky2023, Kumar2023, Raouafi2023}. A detailed study of fine CH structure, including CH boundaries, would shed new light on the relationship between the interior of CHs and embedded solar wind structures, and CH boundary structure and the slow solar wind. 

In this work, we leverage advanced image processing techniques applied to high-resolution and high-cadence extreme ultraviolet images and reveal detailed structuring inside mid-latitude CHs throughout a solar cycle. This allows to present a comprehensive comparison of coronal cells and plumes both in open and closed magnetic field topologies. In Section \ref{sec:data-meth} we describe the data used in this work and the methods used for analyzing the data, in Section \ref{sec:results} we describe in detail our results, in Section \ref{sec:discussion} we discuss our interpretation of the results including our schematic representation of the coronal field composed of cells, and in Section \ref{sec:conclusions} we present our conclusions.

%------------------------------
\section{Data and Methods} 
\label{sec:data-meth}
%------------------------------

This study was carried out using data from the Atmospheric Imaging Assembly instrument \citep[AIA;][]{Lemen2012} onboard the \textit{Solar Dynamics Observatory} (SDO). The AIA instrument images the solar atmosphere in seven EUV channels and three UV channels out to $\sim$1.3 $R_{S}$. For this study, images in the 171, 193, 211 and 304 \AA\ channels were used with a cadence of 12 s when analyzing specific regions and 1 hr for limb to limb analysis. For the three-wavelength (RGB) composite images we use the 171, 193 and 211 \AA\ channels where red is the 171, green is the 193 and blue is the 211 channel. We also used line-of-sight (LOS) magnetograms from the Helioseismic Magnetic Imager \citep[HMI,][]{Scherrer2012}, also onboard SDO\added{, with contrast adjusted via histogram equalization.} The methodology implemented in this work can be separated into three main steps: (i) image processing for fine structure enhancement in EUV observations; (ii) identification of features and properties (i. e., CH boundaries, cell-like structures and lanes, magnetic flux concentrations\added{, cells' lifetime}); (iii) length series spectral analysis for the identification of characteristic scale size of cells.

\begin{figure}[b!]
    \centering
   \includegraphics[width=\textwidth]{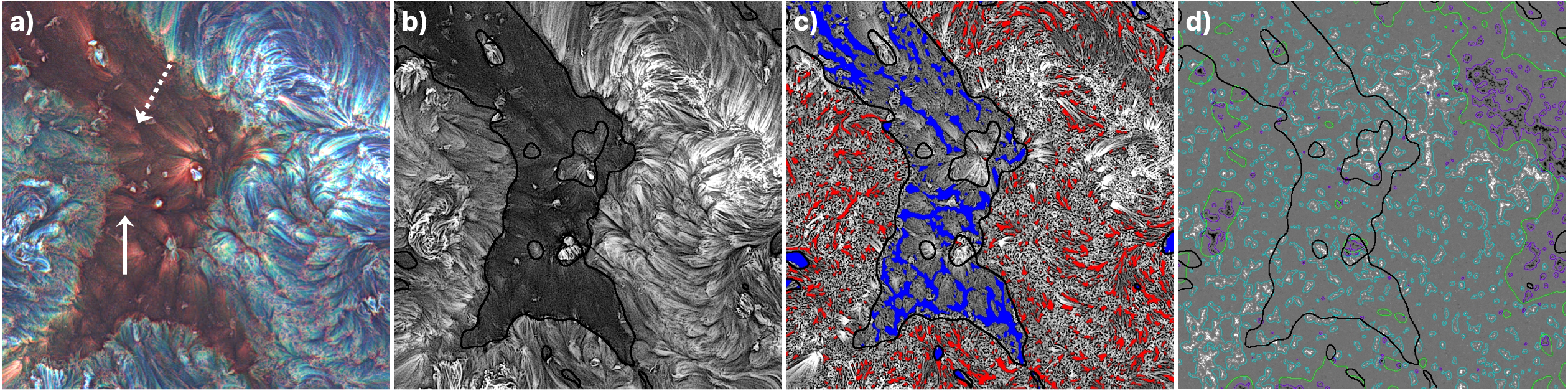}
    \caption{\footnotesize\textbf{Examples of the methods used to identify coronal cells, CH boundary, cells' lanes, and magnetic flux concentrations.} Panel a) shows a large CH in SDO/AIA RGB composite on 31 March 2015; Panel b) shows the same CH but in 193 \AA\ where the CH shows up with better contrast against the surrounding regions; Panel c) shows the 171 \AA\ image plus the CH boundary (black line). Here, the identified coronal cell lanes inside the CH are shown in blue and the outside ones are shown in red; Panel d) shows the outline of the CH overlaid on an HMI magnetogram image showing concentrations of magnetic field flux within the area of coronal cells.}
    \label{fig:contour-steps}
\end{figure}

\subsection{Image Processing for Fine Structure Enhancement in EUV Observations}
\label{sec:img-proc}
%------------------------------

We began by applying a point filtering method to the original unprocessed images to remove outliers and isolated bright spots. We then applied the Multiscale Gaussian Normalization \citep[MGN,][]{Morgan2014} technique to EUV images from the SDO/AIA instrument to reveal the fine structure details inside and outside each CH analyzed. This technique starts with a normalization of the unprocessed EUV image using a Gaussian-weighted sample of local pixels which is then rescaled by an arctan function. This is repeated over several spatial scales. Finally, the weighted combination of the normalized components is added to the original gamma-transformed image. The MGN processed EUV image reveals fine details in the corona and structures in off-limb regions. The image is then restricted to a certain sub-region around the CH of interest to which the image is cropped. Panels c) in Figure \ref{fig:sheeley-compare} and a) in Figure \ref{fig:contour-steps} show a 3-wavelength AIA composite processed with the MGN method showing regions of high contrast. The arrows in Figure \ref{fig:contour-steps}a indicate coronal cell structures inside a CH.

% aims at revealing information that is often hidden in the broad brightness range of EUV images. It normalizes an image by using the local mean and standard deviation calculated using a Gaussian-weighted sample of local pixels. The normalized image is transformed by the arctan function and this process applied over several spatial scales. The final image is a weighted combination of the normalized components, plus the original gamma-transformed image, which reveals fine details in the corona and structures in regions off-limb.

\subsection{Identification of Coronal Cells' Features and Properties}
\label{sec:ID-CHB}
%------------------------------

To identify the CH boundary, we used the 193 \AA\ AIA images since they appear to have greater brightness contrast between the CH regions and the nearby regions. We smoothed the processed images using the IDL Gaussian-Function with $\sigma=20$. The CH boundary is then defined as the contour level corresponding to a certain percentile level of the overall brightness of the cropped AIA 193 \AA\ image. Panel b) in Figure \ref{fig:contour-steps} shows the result for one of our events.

The next step was to identify the coronal cell network and the corresponding lanes. We divided the processed AIA cropped image in two regions, inside and outside the CH boundary as defined above. The two separate regions were further smoothed via the IDL Gaussian-Function with $\sigma=5$ inside the CH and $\sigma=3$ outside the CH. This additional step helps reduce the level of details in the image revealed by the MGN method leading to a more uniform identification of the coronal cell lanes. We defined a certain percentile level of the overall brightness separately for the areas inside and outside the CH boundary below which the coronal cell lanes are defined. Panel c) in Figure \ref{fig:contour-steps} shows the lanes in blue inside the CH and in red outside the CH.

To investigate the connection between magnetic flux concentrations and coronal cells, we identified contour levels of magnetic field flux in HMI data. We smoothed the magnetogram images using the IDL Gaussian-Function with $\sigma=21$ to identify the neutral line (green line in Figure \ref{fig:contour-steps}d). The area of positive and negative fluxes were derived from images smoothed with $\sigma=5$ and defined at levels 10 G (cyan contours in Figure \ref{fig:contour-steps}d) and -10 G (purple contours in Figure \ref{fig:contour-steps}d). \added{Note that, due to the smoothing process, the magnetic field thresholds are not representative of the actual magnetic field strength.  The field might be significantly higher especially in very small areas of flux concentration, that may not even be resolved by the HMI measurements.} Panel d) in Figure \ref{fig:contour-steps} shows the flux concentrations inside and outside the CH outlined with a black line.

\begin{figure}[b!]
    \centering
    \includegraphics[width=0.9\textwidth]{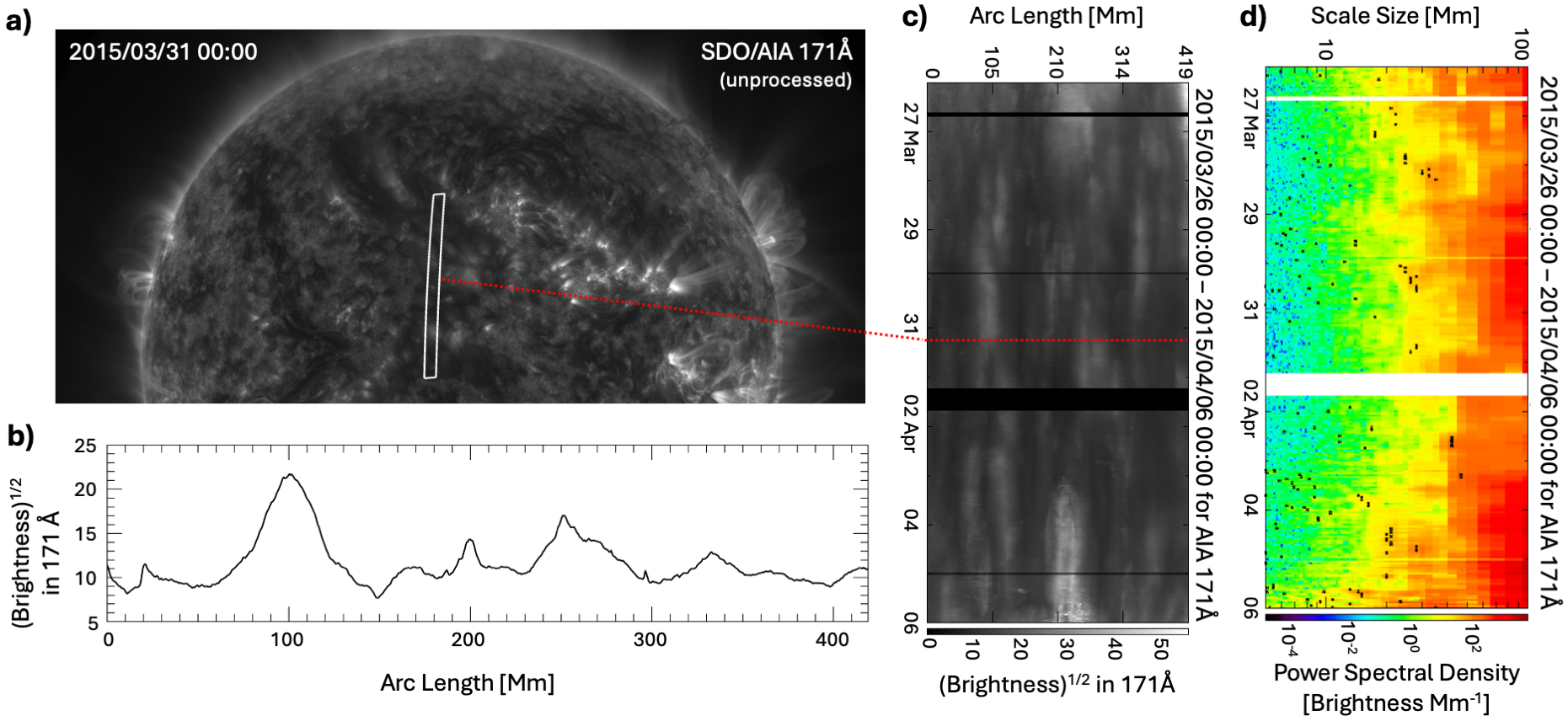}
    \caption{\footnotesize\textbf{Steps of the data preparation and spectral analysis approach used to identify the cells' characteristic size scale.} a) Example of selected slit (white contour) on a SDO/AIA frame; b) Square root of the average brightness profile extracted from the slit; c) Time evolution of the brightness profile obtained as the sun rotates. Black areas indicate missing data; d) Power spectral density (color scales) of the brightness profiles. Black dots identify power enhancements above the 90\% confidence threshold corresponding to the characteristic scale size of observed cells.}
    \label{fig:spectral-meth}
\end{figure}

\subsection{Spectral Analysis of Average Brightness Across Slit}
\label{sec:spectra}
%------------------------------

Figure \ref{fig:spectral-meth}a shows an unprocessed image of SDO/AIA for 2015-03-31 at 00:00 UT. We manually selected a slit (white contour) on the sun surface containing a series of coronal cells. Note that the vertices of the slit are connected by arcs along the sun surface. We then extracted a brightness profile along the longer side of the slit by deprojecting the portion of the image and taking the average brightness values of subareas defined by a uniform step of 0.1$^\circ$ (1.2 Mm) along the arc centered with respect to the shorter side of the slit. Figure \ref{fig:spectral-meth}b shows the square root of the average brightness length series revealing quasi-periodic variations on scales of tens of Mm. By taking into account the sun's differential rotation, we track the same area and extract a brightness profile at each time step for $\approx$11 days. Figure \ref{fig:spectral-meth}c shows the resultant time evolution of the square root of the brightness length series. The black rectangles mark intervals of bad/missing files. To identify the characteristic size scale of the coronal cells, we applied a robust spectral analysis approach \citep{DiMatteo2021}, based on the multitaper method \cite[MTM;][]{Thomson1982} with time-halfbandwidth product $NW$=2.0 and $K$=3 tapers. The software for this approach is open source and available to the community \citep{DiMatteo2020}. This technique enables the separation of the power spectral density (PSD) in a continuous background and discrete power enhancements related to periodic variations. The background estimation is obtained by means of a maximum likelihood fitting of a bending power law function to the PSD. In Figure \ref{fig:spectral-meth}d, Areas of the spectrum above the 90\% confidence threshold, amplitude test, are marked with black dots in the dynamic spectrum and identify a characteristic length scale. Isolated dots are likely related to the expected level of false positives (10\%), while clusters persisting in time constitute the most reliable identifications. In this work we consider the average size scale of each cluster as one identification of the characteristic cells' size scale. Note that the MTM approach provides an additional independent test, the F-test, for the identification of periodicities in a data series which is often combined with the amplitude test to obtain more robust results. In our investigation, the addition of this test reduces the number of selected events but lead to the same conclusions. Therefore, in the following we only show the results stemming from the application of the amplitude test. Applying the same spectral analysis procedure on brightness profiles extracted from the MGN processed images, in which cells are better revealed, we obtain similar results \added{(more examples are available at \url{https://zenodo.org/uploads/15518645})}.

\begin{figure}[b!]
    \centering
    % \begin{interactive}{animation}{figures/Fig4_movie.mp4}
    \includegraphics[width=0.8\linewidth]{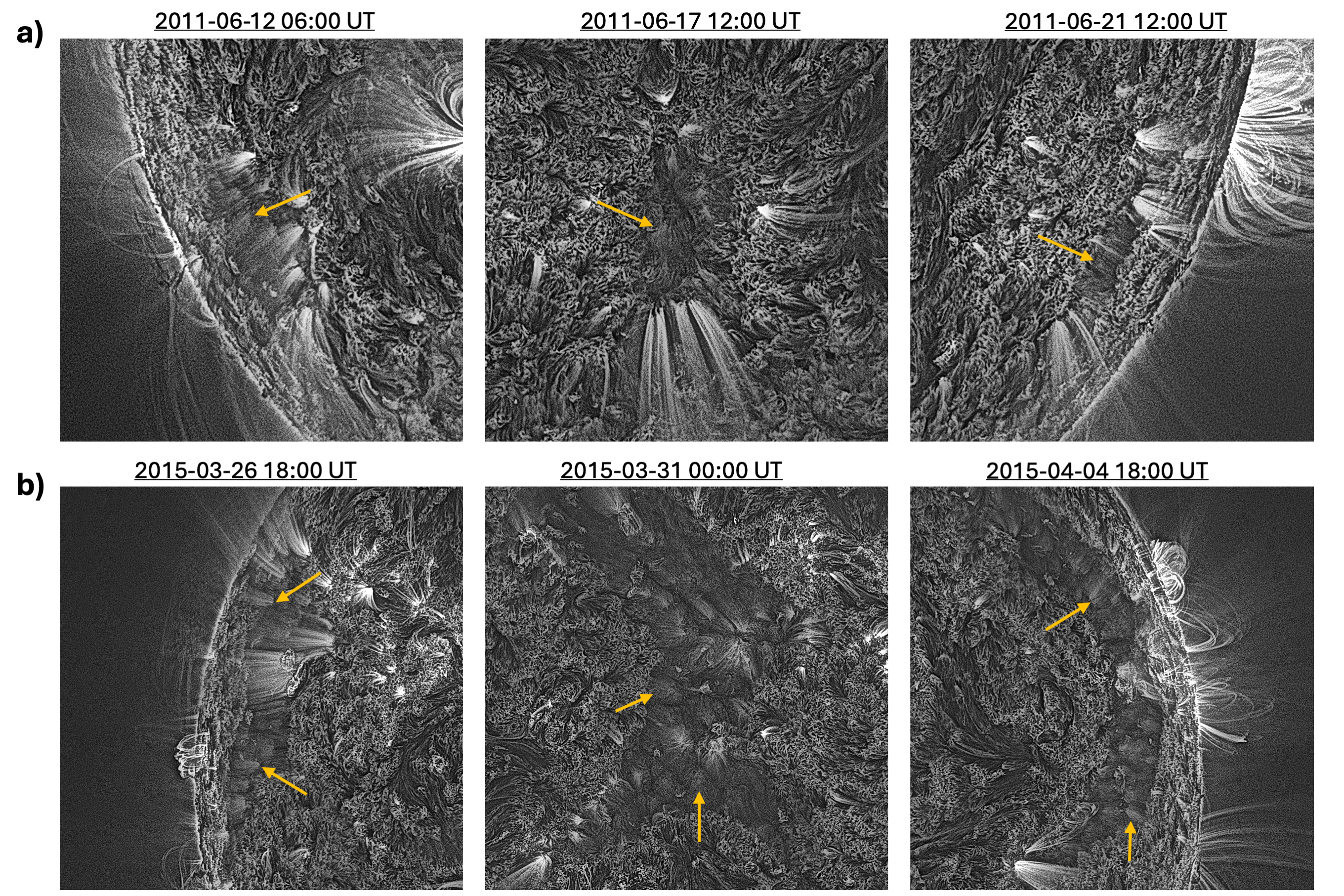}
    % \end{interactive}
    \caption{\footnotesize\textbf{OF-cells compose mid-latitude CH and appear as plumes when observed on limb.} MGN processed SDO/AIA 171 \AA\ observations for the same event analyzed by \citet{sheeley2012} (panel a) and the one in Figure \ref{fig:contour-steps} (panel b) with arrows marking OF-cells. An animation of this figure is available in the online article.  High-resolution versions of the movie in 171, 193, 211 \AA\ and three-wavelength RGB composite at one-hour cadence showing the limb-to-limb coronal hole evolution from 26 March to 06 April 2015 are available at \href{https://doi.org/10.5281/zenodo.15518645}{doi:10.5281/zenodo.15518645}\; \citep{Alzate2025}.}
    \label{fig:disk-limb}
\end{figure}

\begin{figure}[h!]
    \centering
    \includegraphics[width=0.85\textwidth]{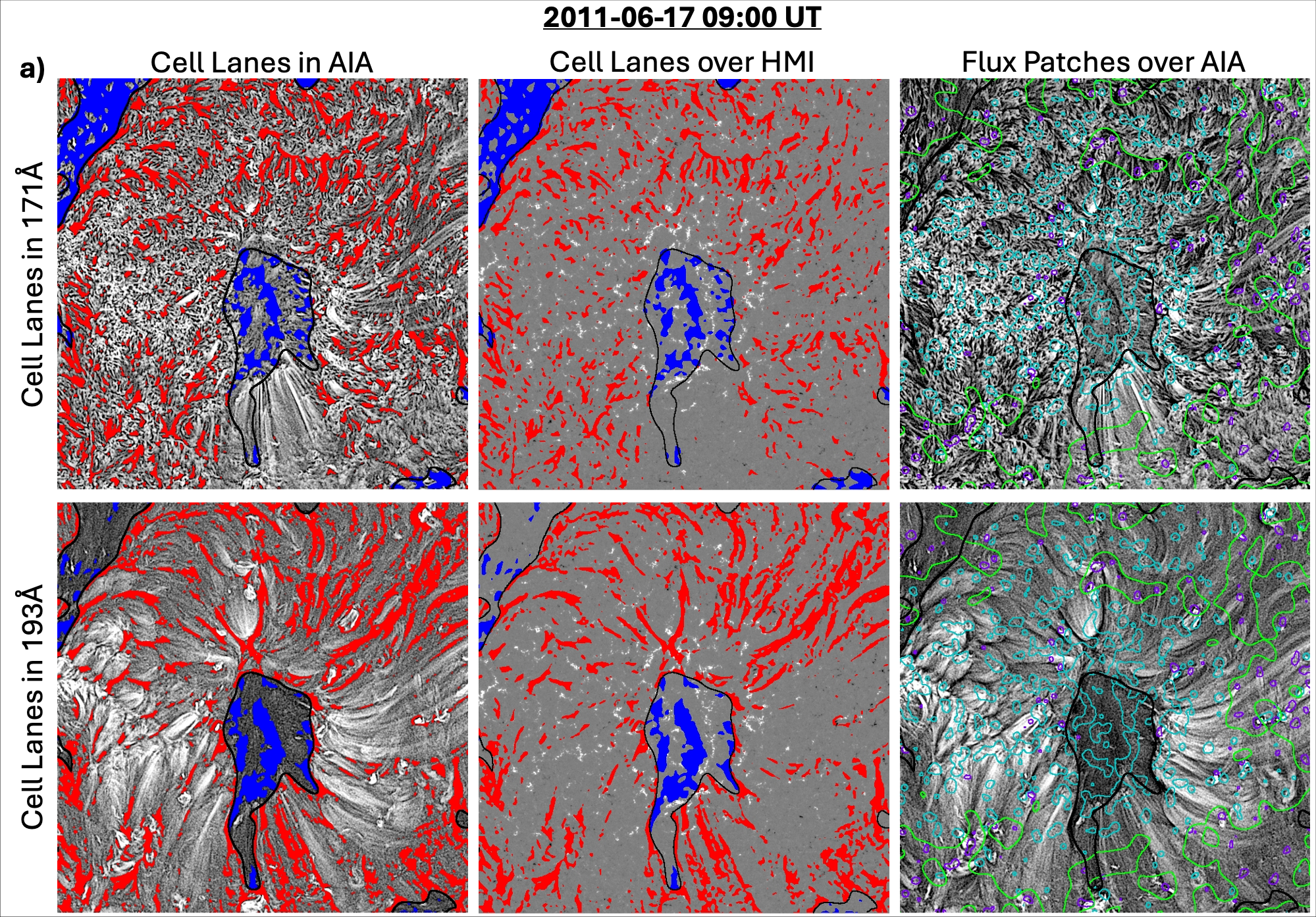}
    \includegraphics[width=0.85\textwidth]{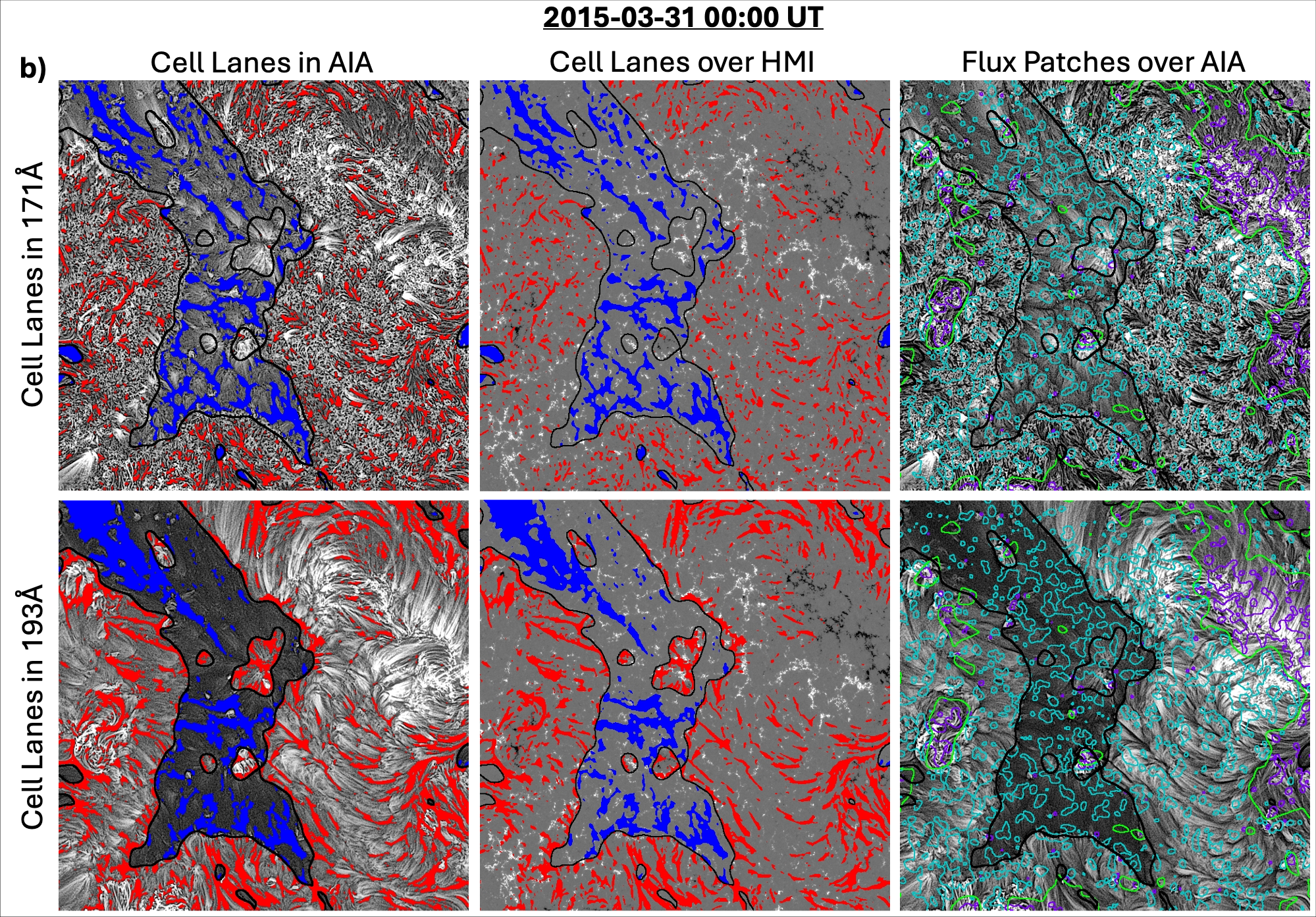}
    \caption{\footnotesize\textbf{Evidence of the existence of coronal cells both inside and outside coronal holes.} The \added{OF}- and CF-cells properties as observed in 171 and 193 \AA\ and HMI magnetogram for the \cite{sheeley2012} event (panel a) and the 31 March 2015 event (panel b). Left column, blue and red identified cell lanes inside and outside each CH, delimited by black lines, as identified in each wavelength. Middle column, the cell lanes are overlaid on HMI magnetograms. Right column, magnetic flux concentration contours above 10 G (cyan) and below -10 G (purple), separated by the neutral line (green), overlayed on AIA images.}
    \label{fig:lanes-results}
\end{figure}

\begin{figure}[h!]
    \centering
    \includegraphics[width=\textwidth]{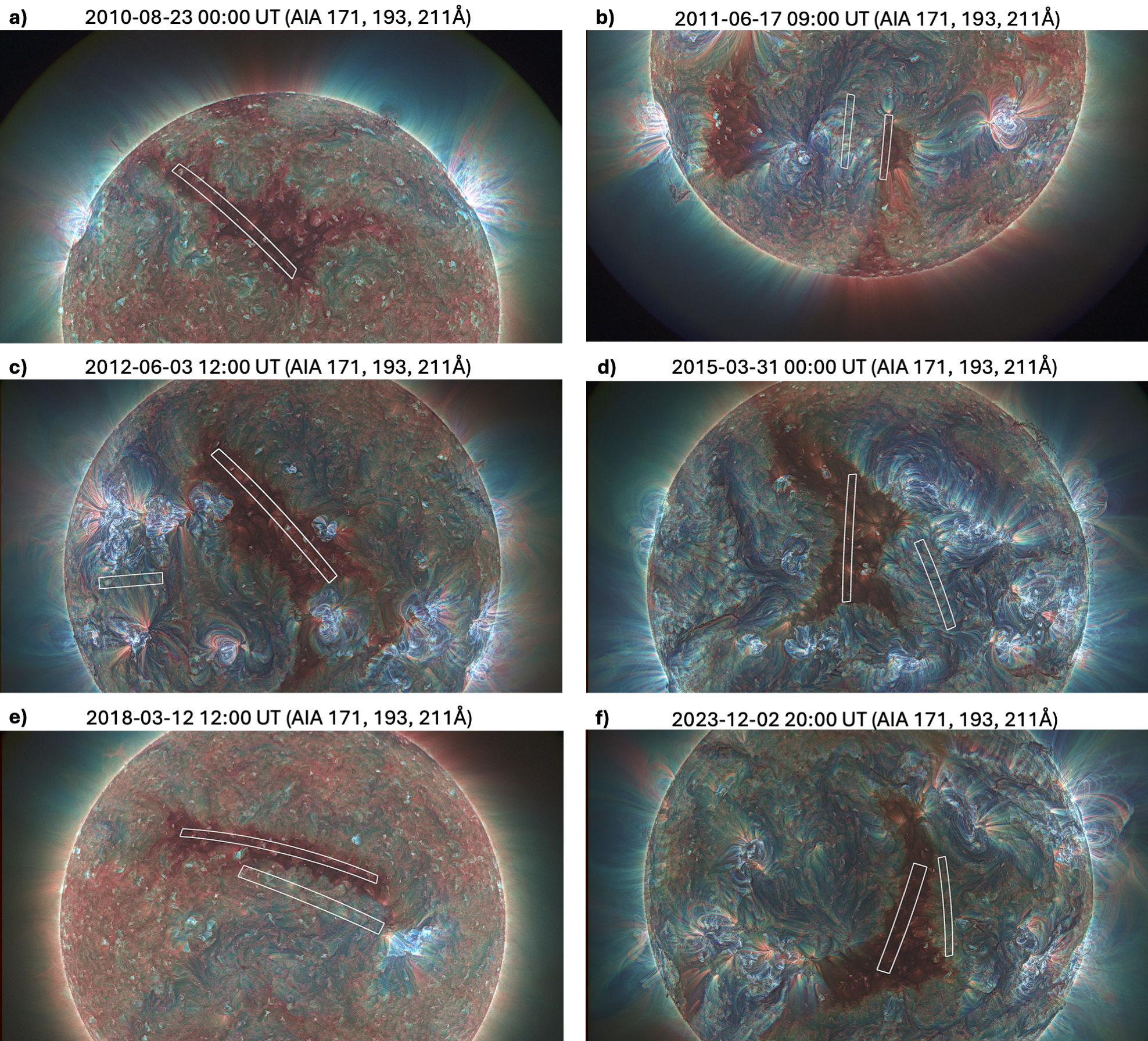}
    \caption{\footnotesize\textbf{Evidence of the existence of both OF- and CF-cells throughout the solar cycle.} SDO/AIA RGB composite showing our target events throughout Solar Cycle 24. White shape indicates slit used for analysis.}
    \label{fig:aia-rgb-sc24-v2}
\end{figure}

%------------------------------
\section{Observations and Results} 
\label{sec:results}
%------------------------------

\added{We first investigated the event from \citet{sheeley2012} in which coronal cells in closed field region (CF-cells) are clearly visible but this time using 171 \AA\ observations. In Figure \ref{fig:disk-limb}a we show MGN processed images centered on the same coronal hole in a face-on view and when near the east and west limb. The middle panel provides proof of the existence of ubiquitous coronal cells inside a CH (Open Field cells, OF-cells). Similar to CF-cells, the OF-cells appear as plumes when observed close to the limb. The arrows indicate examples of OF-cells in each different viewpoint. In Figure \ref{fig:disk-limb}b, we show another example of OF-cells fully composing a larger mid-latitude CH. Note that OF-cells are very faint and their observation can easily be hindered by nearby brighter CF-cells especially close to the limb. Additionally, out of the sun disk at mid-latitudes, where the cells appear as plumes, the overlap of multiple OF- and CF-cells along the field of view results in blurred observations from the mixture of the cells' ray-like features. An accompanying animation is provided for Figure \ref{fig:disk-limb} showing the evolution of OF-cells in SDO/AIA 171 \AA\ for the March 2015 event.}

\added{We also perform a more detailed analysis of the events presented in Figure \ref{fig:disk-limb} focusing on SDO/AIA observations in 171 and 193 \AA\ and the corresponding magnetograms from HMI. Figure \ref{fig:lanes-results} shows that OF}- and CF-cells are similar in shape and size and are characterized by ray-like features. The left column shows that \added{OF}-cells are better identified in 171 \AA\ observations while the CF-cells are clearer in 193 \AA. The cells are delimited by dark lanes both inside (shown in blue) and outside (shown in red) the CH which are best visible in the same wavelength in which the cells are best seen. The HMI magnetograms (middle column) show that the cell lanes, identified in AIA, always surround an area with a small magnetic flux concentration. The polarity of these magnetic flux concentrations is, for the vast majority, the same inside and in proximity of the CH. This is further shown in the third column, in which we compare AIA observations with the magnetic flux concentrations represented by contour levels at +10 G (cyan) and -10 G (purple) separated by the neutral line (green). Note that the magnetic flux concentrations are located at the base of both \added{OF}- and CF-cell, in which the ray-like features of the cells appear to be rooted. There are also some loop-like structures, more visible in 193 \AA\ observations, inside the CH which appear to be associated with bipolar flux at the photosphere. We further support these observations performing a detailed analysis of \added{OF}- and CF-cells for six events throughout Solar Cycle 24. The RGB composite for each event in Figure \ref{fig:aia-rgb-sc24-v2} shows that coronal cells are observed throughout the solar cycle with very similar characteristics. \added{The same analysis, as in Figure \ref{fig:lanes-results}, performed on} each event, \added{available at \href{https://doi.org/10.5281/zenodo.15518645}{doi:10.5281/zenodo.15518645}\; \citep{Alzate2025},} confirmed the persistence of the coronal cells' properties previously discussed throughout the solar cycle.

\begin{figure}[ht!]
    \centering
    \includegraphics[width=0.925\textwidth]{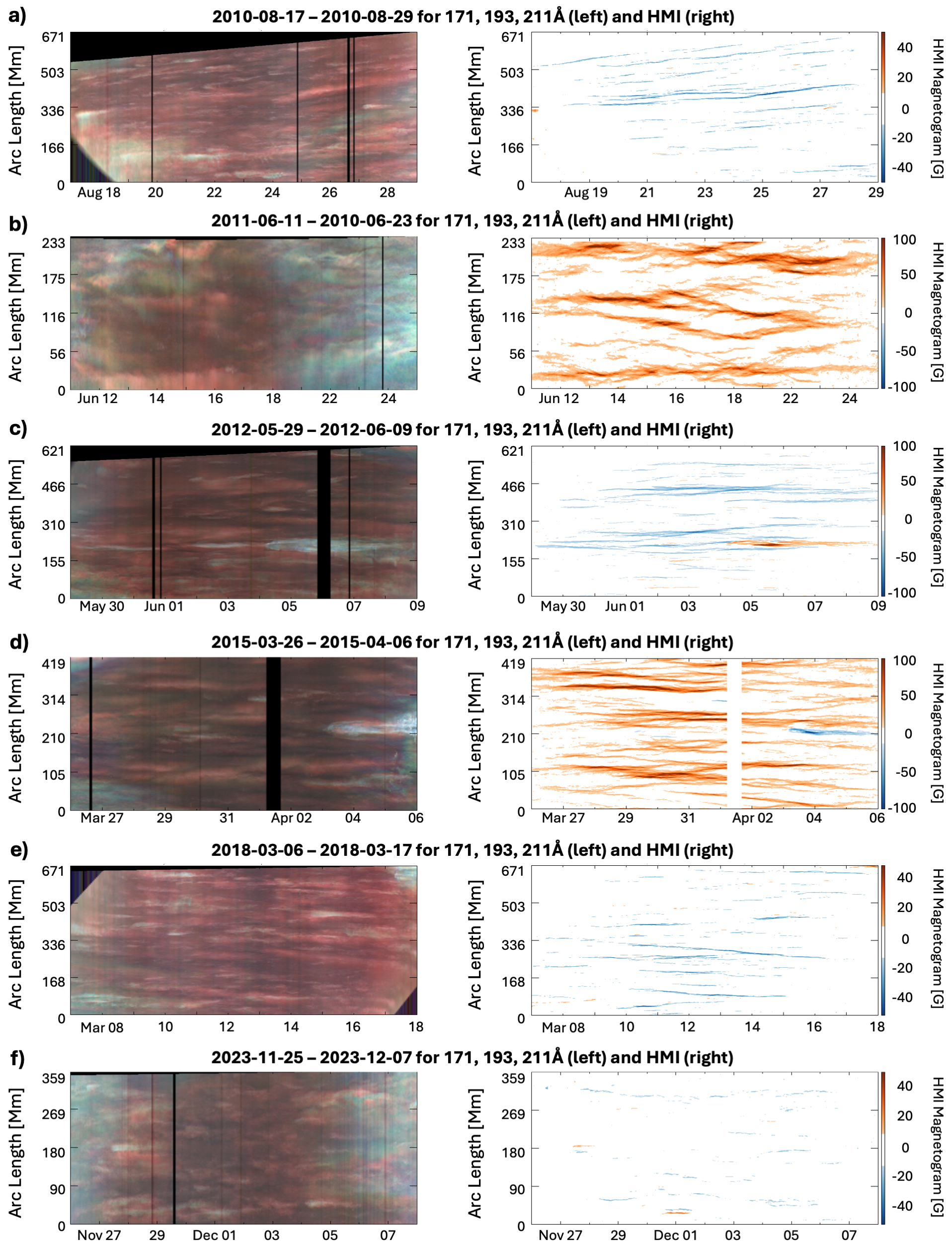}
    \caption{\footnotesize\textbf{The lifetime of OF-cell is determined by the persistence of unipolar magnetic flux concentrations.} In each panel, for \added{OF}-cells, left)  RGB composites of the time evolution of average brightness profile from three wavelengths (171, 193 and 211 \AA) along slits inside CHs as marked in Figure \ref{fig:aia-rgb-sc24-v2}. The black regions indicate data gaps. Right) Time evolution of average LOS magnetic field from SDO/HMI magnetograms along the same slits and capped above 10 G and below -10 G.}
    \label{fig:CH_inside_cell_scale}
\end{figure}

\begin{figure}[ht!]
    \centering
    \includegraphics[width=0.925\textwidth]{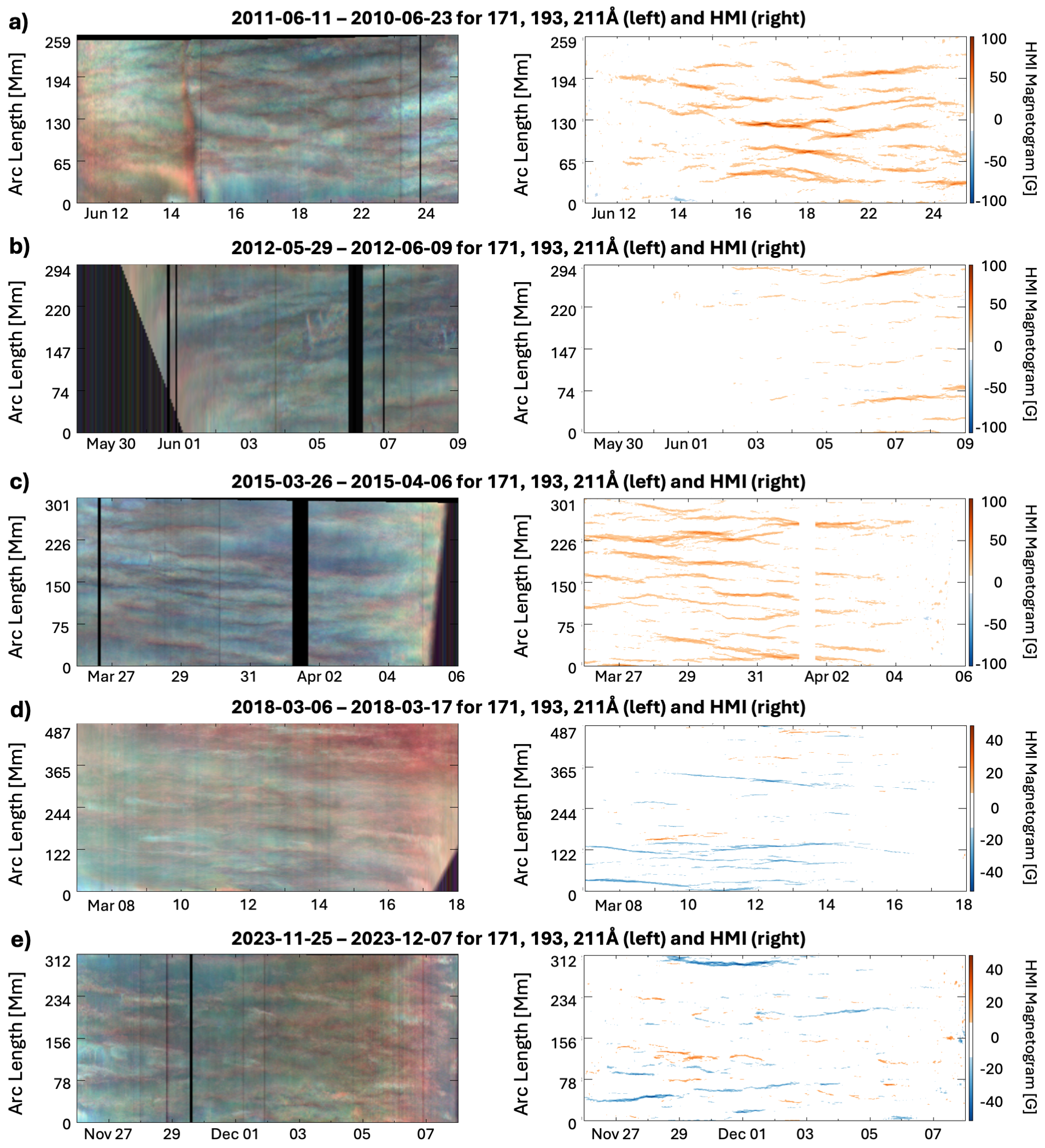}
    \caption{\footnotesize\added{\textbf{The lifetime of CF-cell is determined by the persistence of unipolar magnetic flux concentrations.}} In each panel, CF-cells, left)  RGB composites of the time evolution of average brightness profile from three wavelengths (171, 193 and 211 \AA) along slits inside CHs as marked in Figure \ref{fig:aia-rgb-sc24-v2}. The black regions indicate data gaps. Right) Time evolution of average LOS magnetic field from SDO/HMI magnetograms along the same slits and capped above 10 G and below -10 G.}
    \label{fig:CH_outside_cell_scale}
\end{figure}

\subsection{Lifetime of coronal cells and relation to network magnetic flux concentrations\label{sec:lifetime}}

To estimate the lifetime and size scales of coronal cells, we extracted slits from white rectangles inside and outside each CH as identified in Figure \ref{fig:aia-rgb-sc24-v2}. Note that for the event on 23 august 2010 there were no clear consecutive CF-cells which are usually more evident near an AR that in this case was visible only on the sun limb. Figure \ref{fig:CH_inside_cell_scale} shows the RGB composite of the brightness extracted from the slits inside each CH for a time period of at least 11 days. The vertical black band indicates data gaps. Due to the sun's differential rotation the shape of the slit changes as the sun rotates affecting the actual length of the slit. This effect is reflected in the dark areas at the top and/or bottom of each plot. The \added{OF}-cells appear as bright patches that persist from a few hours to a few days. RGB composites in panels c and d show the emergence of isolated structures (cyan features) with high brightness also in 193 and 211 \AA\ observations. To further investigate the nature \added{of these structures}, we extracted \added{average} values of the HMI magnetograms along the same slits and reported the results on the right of each panel showing only regions with average magnetic field \added{strength} above 10 G \added{(see section \ref{sec:ID-CHB})}. The \added{OF}-cells show a one-to-one correspondence with unipolar flux concentrations of the same polarity dominating the CHs. The brighter structures, also previously reported as coronal plumes \citep{Moore2023}, are associated with stronger magnetic flux concentrations. Lastly, bright structures in 193 and 211 \AA\ wavelength find a one-to-one correspondence with the emergence of bipolar magnetic flux relating them to coronal bright points \citep{Madjarska2019}. Figure \ref{fig:CH_outside_cell_scale} shows the properties of CF-cells \added{whose lifetime also} ranges from a few hours to a few days. Looking at the corresponding HMI magnetograms, on the right of each panel, we can see again how each cell is related to the emergence of unipolar \added{network} magnetic flux concentrations. Also, note that the polarity of these flux concentrations is the same as the one in which the nearby \added{OF}-cells are rooted (except for the event on 3 June 2012, for which the selected CF-cells are far from the CH \added{belonging to a different area of unipolar field concentrations}) indicating that nearby \added{OF}- and CF-cells are part of the same large scale unipolar region. The differences in their properties reside in their magnetic field connectivity. Based on the RGB representation, CF-cells show similarities with the coronal bright points inside CHs (panels c and d in Figure \ref{fig:CH_inside_cell_scale}) associated with closed filed lines. Interestingly, small bipoles emerge also outside the CH forming similar coronal bright points, clearly distinguishable from CF-cells (e. g., panel d and e in Figure \ref{fig:CH_outside_cell_scale}). These observations suggest that cells are a manifestation of unipolar magnetic flux concentrations due to supergranular motion at the photosphere with CF-cell magnetic field closing into areas of magnetic field concentrations of opposite polarity. The \added{OF}-cells appear to be the building blocks of CHs with their magnetic field expanding outward to form the sun open field.

\begin{figure}[b!]
    \centering
    \includegraphics[width=0.6\textwidth]{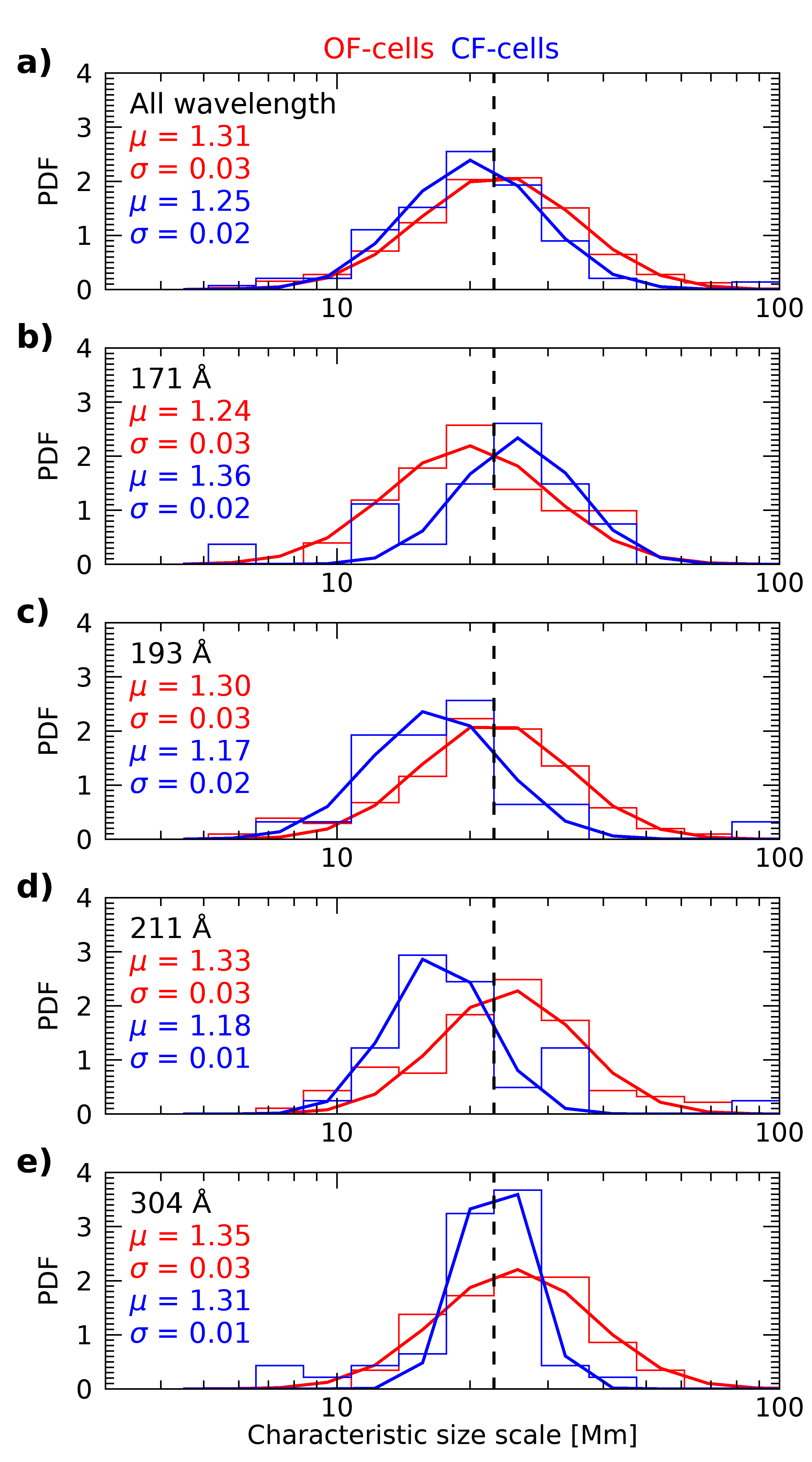}
    \caption{\footnotesize\added{\textbf{The characteristic size scale of both OF- and CF-cells is around the supergranular scale.}} Probability Distribution Functions (PDFs) of the identified characteristic size scales of \added{OF}-cells (red) and CF-cells (blue) for all SC events in this study in all and each wavelength. The vertical lines at 22.6 Mm are used as reference to compare the shift of the PDFs.}
    \label{fig:CH_AR_cell_scale}
\end{figure}

\subsection{Size scales of coronal cells}

Lastly, we investigate the possible relation of cells with supergranules by estimating the cell's size scale. The average brightness length series (see example in section \ref{sec:spectra}) shows quasi-periodic variations. The different amplitude and duration of the brightness variations are related to the characteristic scale of the cells but it could be affected by: (i) the particular choice of the slit, which might include only a portion of some cells; (ii) the lifetime and evolution of cells, which might make them move outside the slit or slowly disappear. We tested the MTM spectral analysis approach \added{(see Section \ref{sec:spectra})} on brightness profiles extracted from various slits. For a poor choice of the slit, not including deliberately the full cells or only isolated ones, we obtained sporadic isolated identification of periodicities like the one in Figure \ref{fig:spectral-meth}d for scales of a few Mm. When we analyze slits containing a series of cells, the spectral analysis reveals a cluster of periodicity identifications persisting in time (see Figure \ref{fig:spectral-meth}d for scales of $\approx$25 Mm). It follows that isolated dots are likely related to the expected level of false positives (10\%), while clusters persisting in time constitute a reliable identification of the characteristic size scale of cells. In this work, we consider the average size scale of each cluster as one identification of the characteristic size scale of cells. This approach effectively reduces arbitrary choices in the definition of the cells' spatial extension and reduces possible sources of uncertainties.

We applied the MTM spectral analysis approach to average brightness profiles extracted from MGN processed images for the six events covering the phases of the solar cycle and for four wavelengths (i. e., 171, 193, 211, and 304 \AA; results for slits extracted from unprocessed and MGN processed 171 and 193 \AA\ images for all events are \added{\citep[][available at \href{https://doi.org/10.5281/zenodo.15518645}{doi:10.5281/zenodo.15518645}]{Alzate2025}}). Figure \ref{fig:CH_AR_cell_scale} shows the total and by wavelength probability density function (PDF) of the characteristic size scales of both \added{OF}-cells (red profiles) and CF-cells (blue profiles; except for the event on 23 August 2010 for which there was no clear series of CF-cells). The total PDF (panel a in Figure \ref{fig:CH_AR_cell_scale}) reveals that the size scales of \added{OF}- and CF-cells follow a log-normal distribution ($Lognormal(\mu,\sigma^2)$). By fitting a normal distribution in logarithmic space, we can estimate the parameters of the distributions which are summarized in \added{Figure \ref{fig:CH_AR_cell_scale}}. The \added{median} of the log-normal distribution is \added{$10^{\mu}\approx 20$ Mm} for \added{OF}-cells and $\approx 18$ Mm for CF-cells. When divided by wavelength, the \added{OF}-cells show size scales close to the values in all wavelengths except for 171 \AA\ for which they appear more often at smaller scales. On the other hand, CF-cells \added{appear} smaller when observed in 193 and 211 \AA\ and larger in 171 \AA. While these results are intriguing, a more extensive analysis is needed to confirm them and draw a robust conclusion.

\begin{figure}[b!]
    \centering
    % \begin{interactive}{animation}{figures/Fig10_movie.mp4}
    \includegraphics[width=\textwidth]{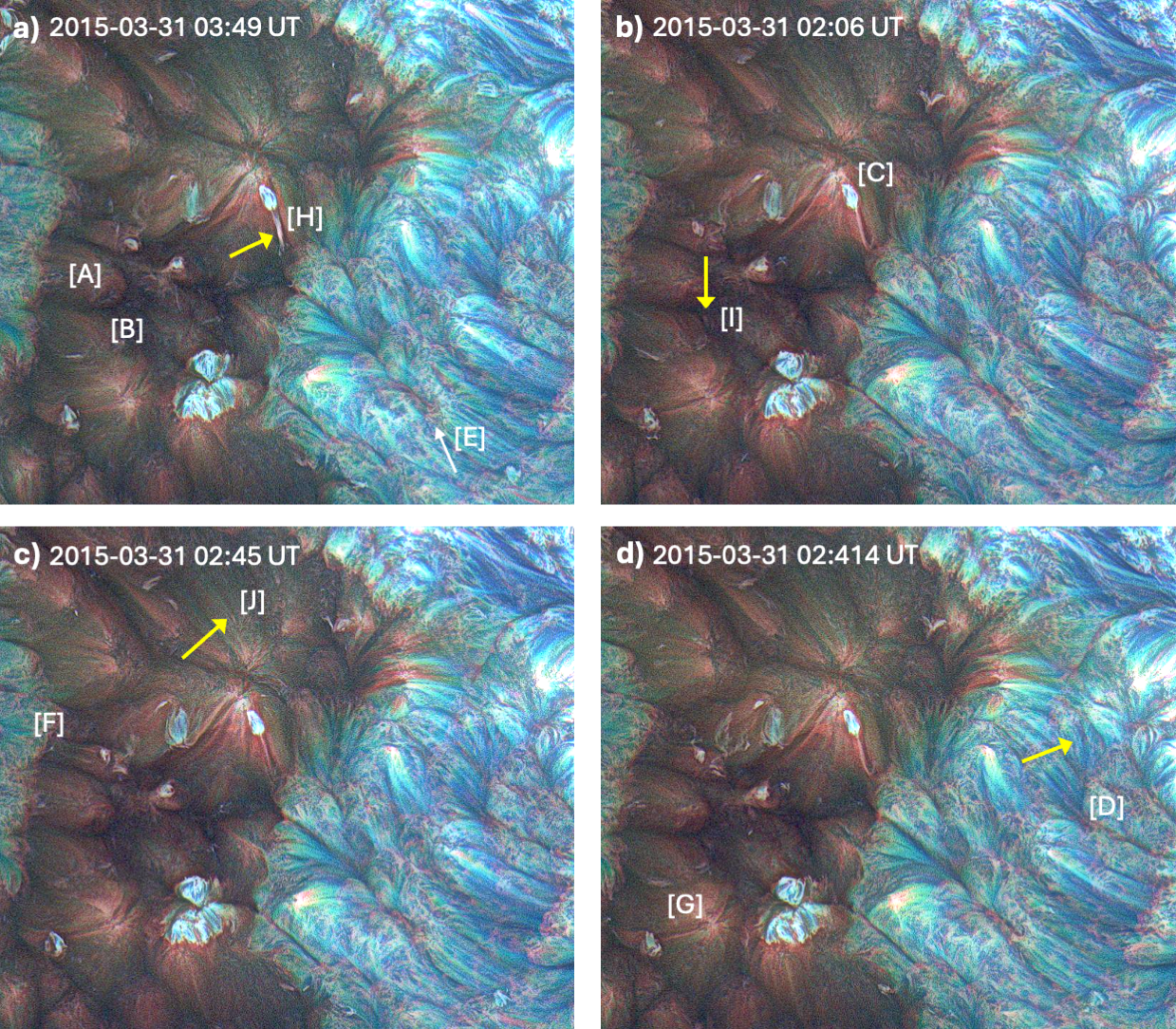}
    % \end{interactive}
    \caption{\footnotesize\added{\textbf{The time evolution of cells reveals ubiquitous small scale jetting activity.}} RGB composites of three wavelengths (171-193-211 \AA). \added{The yellow arrows mark:} a) Jetlet from plume inside a CH. b) Features in dark lanes akin to a pico-flare jet. c) Features along faint ray-like structures, possibly akin to plumelets, consistently moving away from the cell center. d) Features in CF-cells moving along ray-like structures. The letters indicate additional features discussed in section \ref{sec:discussion} and Figure \ref{fig:schematic}. An animation of this figure is available in the online article.  High-resolution versions of the movie in 171, 193, 211 \AA\ and three-wavelength RGB composite at 12-s cadence are available at \href{https://doi.org/10.5281/zenodo.15518645}{doi:10.5281/zenodo.15518645}\; \citep{Alzate2025}.}   
    \label{fig:rgb_movie}
\end{figure}

\subsection{Small scale jetting activity inside coronal cells}

\added{In Figure \ref{fig:rgb_movie} we show an insert of the RGB composite on 31 March 2015 at four different times showing most of the features constantly present throughout the solar cycle providing examples of small-scale activity best observed in the accompanying animation. In panel a, we show a jet releasing plasma (yellow arrow), hotter than the surrounding plasma since it is also visible in 193 and 211 \AA\, along what appears to be a narrow corridor \citep{Raouafi2014,Kumar2022}, supporting that the release of plasma is likely from the nearby coronal bright point. In panel b, we indicate dark lanes which might be due to over-expansion of the magnetic field at the boundary of coronal cells causing the plasma to expand, leading to lower emission. Within dark lanes we observe isolated jet-like ejections \citep[e. g.,][]{Chitta2025}, particularly visible in the area indicated by the arrow. Similarly, but more frequently, we observe ubiquitous outward motion due to jet-like ejections along the ray-like features. Panel c indicates a region and time at which these features are particularly clear. The motion of features is also present in CF-cells \citep{sheeley2012,Morgan2022}, more evident in 193 \AA\ observations (panel d). A movie is provided online showing the various jet-like activities.  High-resolution versions of the movie are available at \href{https://doi.org/10.5281/zenodo.15518645}{doi:10.5281/zenodo.15518645}\; \citep{Alzate2025}. These observations suggest that jet-like activity is potentially ubiquitous and regulated by coronal cells.}

\begin{figure}[ht!]
    \centering
    % \begin{interactive}{animation}{figures/Fig11_movie.mp4}
    \includegraphics[width=0.9\textwidth]{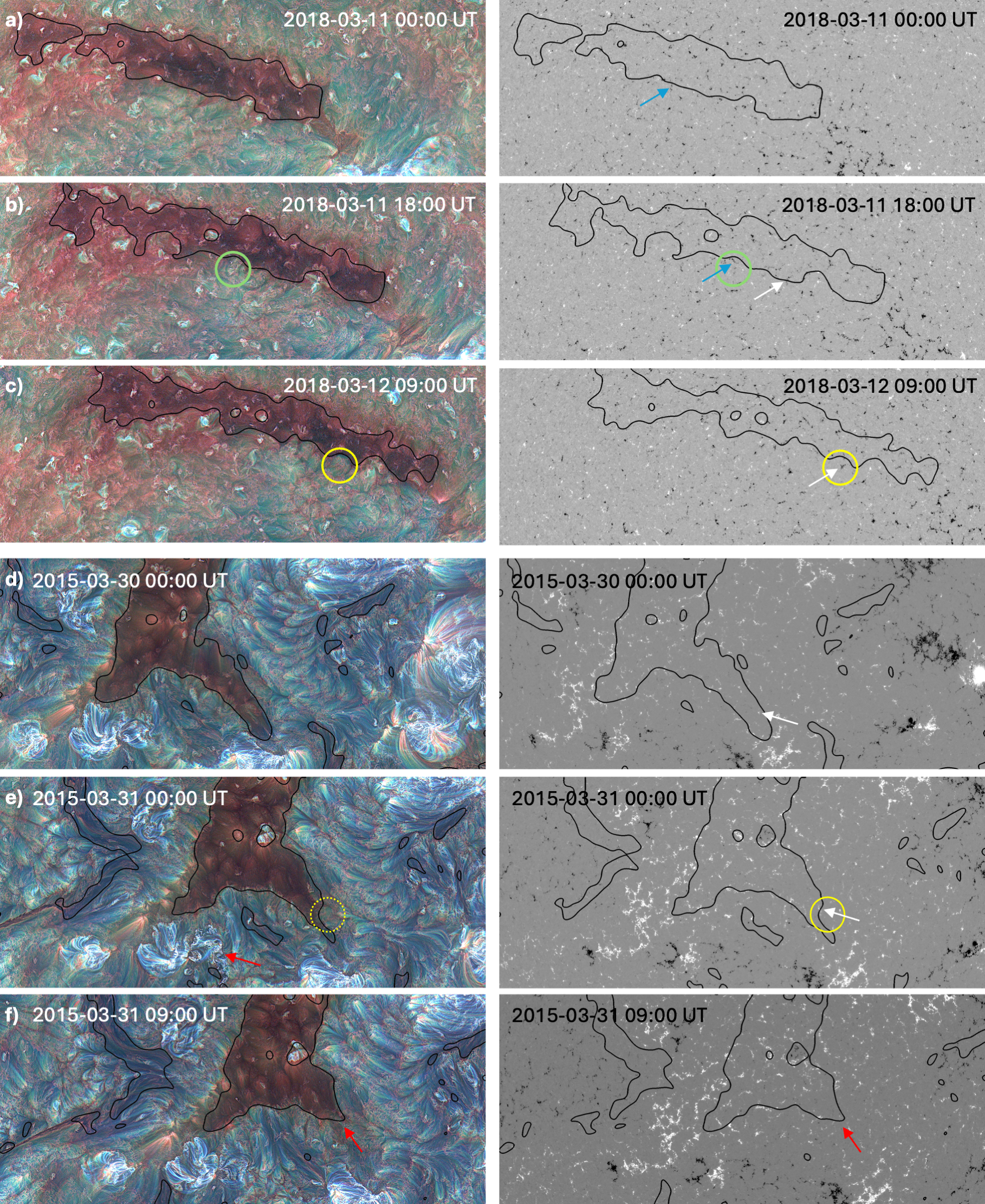}
    % \end{interactive}
    \caption{\footnotesize\added{\textbf{The emergence/decay of a coronal cell modify the CH-boundary on supergranular scales.}} Comparison between SDO/AIA RGB (171, 193 and 211 \AA) composite images and corresponding SDO/HMI frames on 11--12 March 2018 (panels a--c) and 30--31 March 2015 (panels d--f). The emergence of \added{\textbf{CF}}-cells (green and yellow circles) in correspondence with magnetic flux concentrations (blue and white arrow) modify the CH boundary extension. Eruptions from active regions, instead, show considerably larger variations of the CH boundary (red arrows).  \added{An animation of this figure is available in the online article.  High-resolution movies of RGB images and HMI magnetograms at 1-hr cadence are available at \href{https://doi.org/10.5281/zenodo.15518645}{doi:10.5281/zenodo.15518645}\; \citep{Alzate2025}.}}
    \label{fig:chb}
\end{figure}

\subsection{Coronal cells role in CH boundary dynamics}

\added{In our observations, the CH boundary can be defined by the region separating OF- and CF-cells. The RGB composites in Figure \ref{fig:rgb_movie} show that the transition from the open to closed magnetic field regions is marked by an envelope of the dark lanes wrapping all the OF-cells. Given the emergence and decay of a coronal cell, happening over several hours to a few days (see section \ref{sec:lifetime} for cells' lifetime), the CH boundary could be affected over similar time scales.
We investigated this aspect following the evolution of the CH boundary for two time intervals during minimum and maximum solar activity. In Figure \ref{fig:chb}a–c, we compare RGB composites with their corresponding HMI magnetograms on 11-12 March 2018 (solar minimum). The black contour indicates the CH boundary and shows its evolution at three different times. In panel a, the blue arrow indicates an area where unipolar magnetic flux is starting to concentrate along the CH boundary. After $\approx$18 hours, the magnetic flux concentration forms a CF-cell (green circle in panel b) resulting in a fold in the CH-boundary locally reducing its extension. The same process occurred again in the area marked by the white arrow with another CF-cell (yellow circle in panel c) forming in about 15 hours. Note that the CH-boundary evolution appears to also be affected by the differential rotation and the emergence/decay of bipolar structures \citep{Wang1996,Subramanian2010}. Coronal cells can affect the CH-boundary dynamics during active periods as well. The white arrow in panels d–e mark a unipolar magnetic flux concentration forming a CF-cell (yellow circle) and a fold in the CH boundary in about 24 hours.
Magnetic flux eruptions can also modify the extent of the CH \citep[e. g.,][]{Sheeley1989,Gutierrez2013}. However, during our two examples, large eruptions were absent. Additionally, this process is substantially different from the one related to the emergence and decay of coronal cells. After the emergence of the CF-cell shown in panel e, an eruption occurred in a nearby active region (red arrow). As a consequence, the bottom right leg of the CH suddenly shrinks in about 9 hours (red arrow in panel f). In this process, the CH boundary is modified over several supergranular size scales, with the eruption essentially changing the magnetic connectivity of already existing coronal cells. Our observations show a different process in which the coronal cell's emergence or decay can create folds on the CH boundary on the order of a supergranular scale unit on timescale of a few hours to a few days. This can also be better seen in the accompanying animation of Figure \ref{fig:chb}.
While we showed examples in which coronal cells appear to influence the extent of the CH boundary, more extensive analysis are needed to fully understand their dynamics, impact and connectivity.}

%------------------------------
\section{Discussion} 
\label{sec:discussion}
%------------------------------

\begin{figure}[b!]
    \centering
    \includegraphics[width=\textwidth]{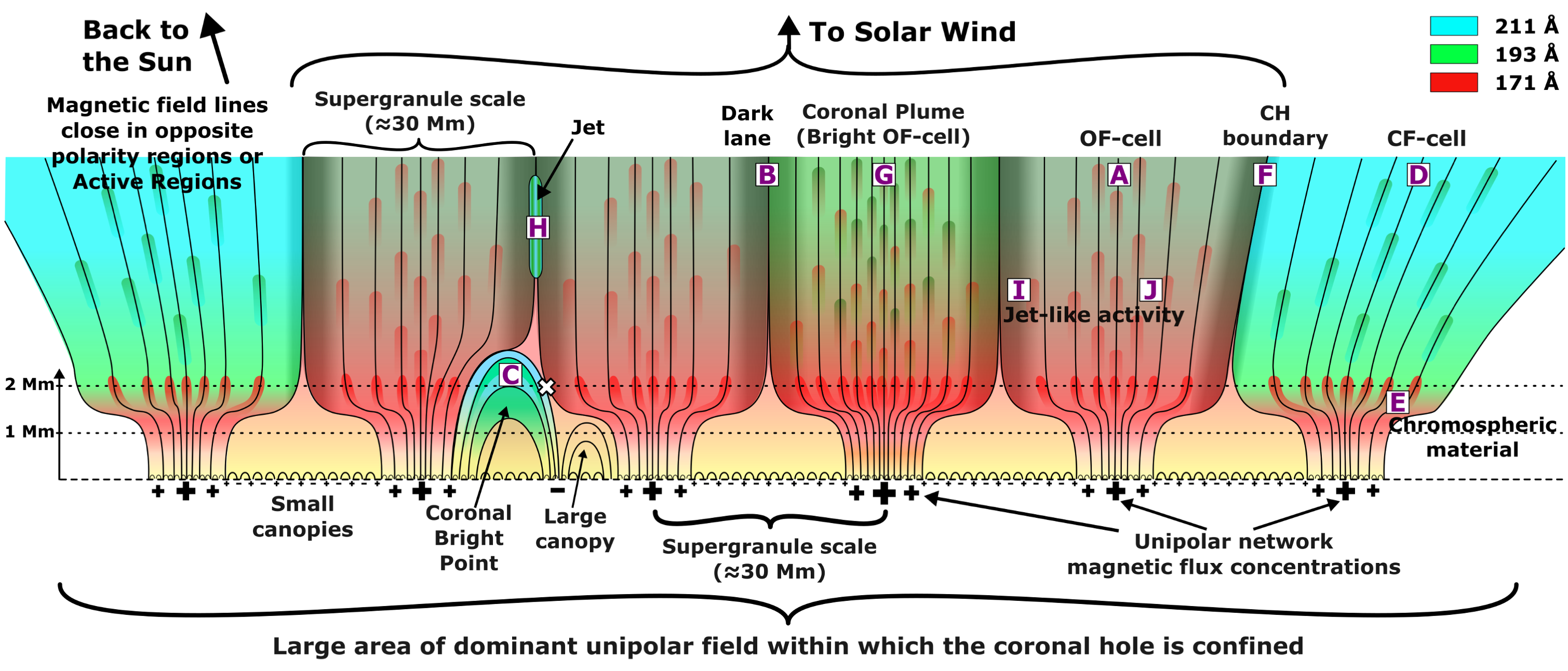}
    \caption{\footnotesize\textbf{Schematic representation of the ``coronal cells'' theory.}  \added{High-resolution full schematic representation and single relevant elements are available at \href{https://doi.org/10.5281/zenodo.15518645}{doi:10.5281/zenodo.15518645}\; \citep{Alzate2025}.}}   
    \label{fig:schematic}
\end{figure}

Our observations show that cells are an ubiquitous consequence of \added{network} magnetic flux concentrations within a large unipolar area, both inside and outside CHs. In Figure \ref{fig:schematic} we present a schematic of the magnetic field and plasma conditions which help explain our observations, expanding on previously suggested magnetic furnace and coronal funnel models \citep{McKenzie1998,Tu2005}. \added{We use} capital letters \added{to} relate features \added{observed in Figure \ref{fig:rgb_movie}} with their schematization in our interpretation. Similar to the RGB composite, we indicate with red, green, and blue a very simplified distribution of the regions from which observations in 171, 193, and 211 \AA\ come from, respectively. The CH volume can be separated into different regions: (i) \added{OF}-cells [A], (ii) dark lanes delimiting \added{OF}-cells [B], (iii) occasional coronal bright points [C]. When the magnetic field funnel closes back to the sun in opposite polarity regions, we obtain hotter regions forming CF-cells [D]. These cells are not visible in 171 \AA\ observations which instead show underlying isolated extensions apparently stemming from the magnetic flux concentration, which \added{might be} related to underlying \added{chromospheric material} [E]. \added{Note that our schematization of the region between the photosphere and chromosphere is a very simplified version of more detailed ones \citep[e. g.,][]{Wedemeyer2009}.} The region between \added{OF}-cells and CF-cells defines the CH boundary [F]. Inside a CH, cells appear with different levels of brightness. The more concentrated the magnetic flux is, the brighter the structure appears, both at the base and up in the corona. Often, the bright structures [G] have been classified as coronal plumes \citep{Moore2023}. \added{We also demonstrated that OF-cells, like CF-cells, appear as plumes when observed close to the limb. This suggests that the} ubiquitous presence of \added{OF}-cells \added{might extend} to polar CHs as well. Interestingly, our distinction between faint cells and bright cells is consistent with previous work which refers to similar structures as ``network plumes'' and ``beam plumes'' \citep{Gabriel2009}. It may be that just as cold plasma suspended in the corona is referred to as a ``filament'' when observed on disk and a ``prominence'' when observed off-limb, these \added{OF}-cells are in fact the on-disk signature of the same magnetic structure making up the polar network and beam plumes. In our picture, the schematic in Figure \ref{fig:schematic} is actually a view of the polar cells from Earth. However, multiple cells overlap along the field of view, result in a blurred view which is a mixture of ray-like features of different colors in our RGB composite that sometimes \added{might} enable the distinction between \added{OF}-cells and CF-cells (see for example the poles in panel c of Figure \ref{fig:sheeley-compare} as well as panel a and b of Figure \ref{fig:aia-rgb-sc24-v2}).  While \cite{Gabriel2009} proposed plume-like extensions along the network lanes to form ``curtains'' to explain these observations, we suggest there is a presence of cells in magnetic flux concentrations along the \added{network} lanes. If this is confirmed, the views of the sun pole would reveal the same cellular structures we are observing in mid-latitude CHs, but with dimensions and lifetimes reflecting the photosphere convection pattern at the sun poles. This prediction could be confirmed by pole observations by Solar Orbiter or by future missions dedicated to the observation of the sun poles \citep{Decadal2024}.

\added{Our observations also provide examples of small-scale activity from jets [H] and jet-like activity along both dark lanes [I] and the cells ray-like features [J]. In our schematic, the jet might originate from the interchange reconnection (indicated by a white cross) of open field lines of an OF-cell and closed field line of a coronal bright point which then releases closed field plasma. The smaller scale activity can be tentatively put in the context of previous works focused on coronal plumes:} the jet-like activity would be the jetlets \citep{Raouafi2014,Kumar2022,Chitta2025}, and our ray-like features would be plumelets \citep{Uritsky2021}. \added{However, note that now these elements would belong to all OF-cells with} jets not required as a precursor to see similar activity. \added{In a proposed scenario in which the solar wind originates from small scale jetting activity in plume like features \citep{Raouafi2023}, previous} analysis found it difficult to reconcile how coronal plumes, estimated to occupy about 10\% of a CH’s volume \citep{Ahmad1977}, could sustain the solar wind emanating from the entire CH. Our new observations show that cells, which intrinsically include coronal plumes, extend to the entire CH, are always present throughout the solar cycle, and therefore can, in principle, sustain the entire CH volume.

\added{Finally, we showed how the emergence and decay of coronal cells changes mid-latitudes CH boundaries on size scale of the order of a supergranule (cells' characteristic size scale) in a few hours to a few days (cells' lifetime). While the detailed dynamics by which this happens will be subject of a future investigation, we remark how the understanding of the CH boundary dynamics on the size scale of supergranules impact many aspects of solar physics. For example, in the interaction of large scale closed field regions (e. g., helmet streamers) with the open field regions from CHs, a processes often invoked at the open-closed boundary to help explain the origin of the slow solar wind is interchange reconnection driven by supergranular photospheric motion \citep[e. g.,][]{Higginson2017}. Another example is strong magnetic fields in mid-latitude CHs and in close proximity to ARs that when mapped on grid sizes on the order of supergranules appear as the primary source of missing open flux and, therefore, as a resolution for the open-flux problem. \citep{Arge2024}. By considering the OF-cells as building blocks of CHs and extending the knowledge of CF-cells properties to OF-cells, this ``coronal cell'' theory might shed new light on these and many other phenomena.}

%------------------------------
\subsection{Possible implications of the ``coronal cell'' theory} 
\label{sec:conclusions}
%------------------------------

\added{In this section, we aim to illustrate possible interpretation of some aspects of our observations, that are still elusive, but provide direction for future investigations.

Each OF-cell might be considered a unit of the open field region giving rise to the solar wind. However, following our schematic, the formation of the solar wind might be more variegated. At least five kinds of streams might stem from: (i) a dark lane between two faint OF-cells; (ii) a dark lane between a faint and a bright OF-cell; (iii) a dark lane in which closed field plasma from a coronal bright point could be released forming a jet; a dark lane forming the CH boundary adjacent to a CF-cell closing (iv) into nearby or far opposite polarity magnetic flux concentrations or (v) nearby active regions. The low emission from dark lanes might be a consequence of a greater radial expansion of magnetic field line with respect to the ray-like features within cells, which might determine lower coronal density and slower solar wind speeds from CHs. Two different kinds of streams might also originate from within faint or bright OF-cells, respectively, where the jet-like activity also appears to be higher possibly providing additional heating, hence possible higher solar wind speeds \citep{Withbroe1988,Poletto2015}. Further investigations might be able to show how different kinds of solar wind from CHs \citep[e. g.,][]{DAmicis2025,Huang2025} might be related to the variety of formation regions introduced by the presence of OF-cells.

Our observations show that the base of CF-cells is likely composed of cooler chromospheric material. If OF- and CF-cells are manifestations of the same phenomena distinguished by only their magnetic connectivity in the corona, then it is likely that the same chromospheric material characterizes OF-cells as well. This is difficult to see with our observations since both the base and the ray-like features of OF-cells are in 171 \AA\ observations. Interestingly, \citet{Samanta2019} showed that chromospheric material, in the form of spicular activity, is associated with jet-like features due to magnetic reconnection at network boundaries. They also showed that even though small flux emergence is ubiquitous in the quiet sun, only the small bipolar flux concentrations close to the strong network fields generate spicules. However, previous reports of dissimilarities between spicules and coronal jets pose questions about this interpretation. In fact, reported correlation between spicules and EUV observations is limited to bright coronal plumes during quasiperiodic emergence of the minor polarity concentrations \citep{Lee2024,Nived2022}. This quasi-periodic dynamics might play a role in the distinction of a general OF-cell and brighter OF-cells.

Noteworthy, recent works also suggested a possible link between magnetic reconnection at magnetic flux concentrations along the network lanes (i. e., the base of coronal cells) and spicular activity to microstreams/switchbacks and additional heating \citep{Samanta2019,Lee2024}. Additional works highlighted how switchback patches, when back-mapped to the sun, correspond to supergranular scales \citep{Fargette2021,Bale2023}. This might motivate possible investigations relating these phenomena to our OF-cells. 
}

%------------------------------
\section{Conclusions} 
\label{sec:conclusions}
%------------------------------

In this work, we provide the first morphological overview of coronal cells as an ubiquitous and persistent consequence of \added{network} magnetic flux concentrations. \added{As such, the presence of cells have been extended inside CHs setting the basis of a ``coronal cell'' theory. Its} key points are as follows:

\begin{itemize}
    \item \added{Both cells inside (OF-cells) and outside (CF-cells) mid-latitude CHs appear as cells \added{confined by dark lanes} when viewed on disk and as plumes when viewed close to sun limb.}
    
    \item The cells' lifetime ranges from a few hours to a few days, emerging and \added{decaying} together with their corresponding magnetic flux concentration in the photosphere. The cells have size scales on the order of super-granules and are always present over the course of a solar cycle.
    
    \item  The cells' characteristics are related to properties of the photospheric magnetic flux in which they are rooted and the subsequent magnetic connectivity in the corona. \added{The OF-}cells extend to form the open field \added{with} their brightness, \added{mainly} in the 171 \AA\ wavelength, \added{possibly} related to the intensity of the magnetic flux concentration \added{or} the presence of quasi-periodic \added{dynamics}. \added{The CF-}cells are among the components of the closed field and their brightness is mostly in the 193 \AA\ wavelength.
    
    \item  Outward propagating jet-like features are present inside the CH and can be separated into \added{three} categories depending on their source region: \added{i)} classic jets from bright \added{OF-cells and/or coronal bright points; ii) persistent} jet-like activity stemming from \added{the cell's base with correspondence to the network field concentration; and iii) occasional jet-like activity in dark lanes.}
\end{itemize}

By quantifying the characteristics of coronal cells and organizing them in the context of previous observations (Sheeley coronal cells, coronal/EUV plumes, ``network plumes'', etc.), we believe our schematic provides a unifying picture of many of these elements. By articulating the possible links of cells down to the chromosphere and up to the open/close coronal magnetic field, \added{further investigations in the context of this ``coronal cells'' theory might be able to provide an explanation} for fine scale coronal structures and heating, which subsequently has a direct effect on solar wind origin.

\begin{acknowledgments}

N.A. acknowledges support from NASA ROSES through HGI grant No. 80NSSC20K1070 and PSP-GI grant No. 80NSSC21K1945. The work of S.D. was supported under the PSP-GI grant No. 80NSSC21K1945.  A. H. was supported in part by the NASA HSOC Program. The authors would like to thank the anonymous reviewer for providing constructive feedback which has significantly improved the clarity of this work.

\end{acknowledgments}

\begin{contribution}
%%This section gives authors the space to recognize author contributions. The text inside this environment is NOT counted towards the total word quanta. At a minimum, manuscripts are expected to include this text:

N.A. and A.H. formulated the original idea of this work. N.A. performed the processing and analysis of AIA and HMI images and extracted the properties of CH boundaries, cell lanes, and magnetic flux concentrations. S.D. carried out the slits analysis for AIA and HMI images and subsequent spectral analysis. All authors participated in the formulation of the final schematic and contributed equally in writing and finalizing the manuscript.

%% But authors are expected to provide more specific details, e.g. 
%%
%%SC was responsible for writing and submitting the manuscript.
%%WWM came up with the initial research concept and edited the manuscript.
%%OTS obtained the funding and edited the manuscript.
%%EBF provided the formal analysis and validation. He also edited the manuscript.
%%GEH Supervised the undergraduates, wrote the software and administers the project github and Zenodo repositories.
%%
%% Authors can use the Contributor Role Taxonomy (CRediT) at
%% for ideas on how write a good statement tailored to their needs.

\end{contribution}

%% in the research for the paper. Each keyword is check against the master 
%% list during copy editing. Individual instruments can be provided in 
\facilities{SDO (AIA and HMI)}

\software{The MGN code is available at \url{https://solarphysics.aber.ac.uk/Data.html}. The spectral analysis code used in this work is freely available on the Zenodo platform \citep{DiMatteo2020}}


\bibliography{coronal_cells}{}
\bibliographystyle{aasjournalv7}

\end{document}